\documentclass[12pt,a4paper]{article}
\usepackage{amsfonts}
\usepackage{color,caption}
\usepackage{graphicx}
\usepackage{amsmath}
\usepackage{float}
\usepackage{amssymb}

\begin{document}

\begin{center}

{\Large \bf Fast scrambling as Brownian motion in a fluid\\ with negative viscosity} \\
\vskip .5cm
\bigskip\bigskip
\bigskip\bigskip

K. Ropotenko\\
\centerline{\it Taras Shevchenko National University of Kyiv}
\centerline{\it 4g Academician Hlushkov Ave.,  Kyiv, 03022, Ukraine}
\bigskip
\verb"ropotenko@ukr.net"

\end{center}
\bigskip\bigskip
\begin{abstract}
It is shown that the fast scrambling of information in a black hole
can be viewed as Brownian motion of information in a fluid with
negative viscosity (and negative temperature). It is argued that a
non-local character of the fast scrambling is only an illusion; the
stretched horizon with negative viscosity is an amplifying medium
that mimics non-locality and superluminal communication.

\end{abstract}
\bigskip\bigskip
\vskip 2cm

\centerline{\it"Essay written for the Gravity Research Foundation
2017 Awards for Essays on Gravitation"} \centerline{(submitted on
March 31, 2017)} \vspace{9cm}

\bigskip\bigskip\bigskip\bigskip\bigskip\bigskip\bigskip\bigskip
\bigskip\bigskip\bigskip\bigskip\bigskip\bigskip\bigskip\bigskip

There are two fundamental problems in black hole physics:

1. What are the degrees of freedom (d.o.f.) responsible for the
Bekenstein-Hawking black hole entropy?

2. How does they process quantum information?

Usually these problems are discussed separately in the literature.
But they are closely related; knowing the d.o.f. we can answer the
second question and vice versa. The Bekenstein-Hawking entropy is a
concept defined in the rest frame of an external fiducial (fixed
$r$) observer. From the viewpoint of the observer, the black hole's
d.o.f. reside on the stretched horizon, a two-dimensional timelike
surface ('membrane') located roughly one Planck length away from the
event horizon and endowed with certain mechanical, electrical and
thermal properties. The number of the d.o.f. is of order the area of
the horizon in Planck units. These d.o.f. absorb, thermalize and
re-emits all information in the form of Hawking radiation. In
particular, from the point of view of the fiducial observer, a
charge falling into the horizon spreads over the entire horizon.
This is in contrast with the point of view of a free infalling
observer who does not experience the horizon. The principle of the
black hole complementarity states, however, that there is no way to
compare observations inside and outside the horizon to find a
violation of the quantum no-cloning theorem.

From the viewpoint of the fiducial observer, the black hole's d.o.f.
thermalize quantum information in the same way as the charge. The
thermalization or scrambling time $t_{\rm scr}$ is one of the most
important parameters of the process \cite{sus1}, \cite{sus2}. For a
system with $N$ d.o.f., it is a measure of how long it takes for
information about a small $O(1)$ perturbation in a system to spread
over the $O(N)$ system's d.o.f. The crucial importance of the
scrambling time consists in the fact that knowing it we can get
insights into the nature of the black hole's d.o.f. responsible for
the Bekenstein-Hawking entropy and into the mechanism of information
retrieval from evaporating black holes. Can this time be made
arbitrary small?

In \cite{hay}, Hayden and Preskill found that for a black hole with
temperature $T$ and entropy $S$
\begin{equation}
\label{1} t_{\rm scr} T \sim \log S,
\end{equation}
and argued that we must put a lower bound on the scrambling time to
avoid a violation of the quantum no-cloning theorem. Therefore, $
t_{\rm scr} T \geq \log S $. However, this is a much shorter time
scale as a function of the entropy than the time that a Brownian
particle would take to diffuse over the entire horizon. Indeed, by
using diffusion approximation, Seiko and Susskind \cite{sus1}, and
also Susskind \cite{sus2}, showed that for ordinary local systems,
such as quantum field theories or fluids,
\begin{equation}
\label{2} t_{\rm scr} T \sim S^{\frac{2}{d}},
\end{equation}
where $d$ is the number of spatial dimensions. Comparing (\ref{1})
and (\ref{2}), Seiko and Susskind conjectured that black holes are
the fastest scramblers in the nature. The stretched horizon is a
two-dimensional system. But if measured by its scrambling time it is
more like infinite dimensional systems. Thus, Seiko and Susskind
concluded that the fast scrambling is a result of the non-local
interactions between the black hole's d.o.f.; local interactions, in
their opinion, would lead to slow diffusion as in Brownian motion.

In this essay we shall show that this conclusion is not quite true.
The point is that it is valid only for fluids with positive
viscosity. However, as Damour showed \cite{dam}, the stretched
horizon behaves as a fluid with a \emph{negative viscosity}. As a
result, a Brownian particle moving in such a fluid will not slow
down, but accelerate; as will be demonstrated below, the mean-square
displacement of the particle grows exponentially with time. By using
this fact, we shall show that the fast scrambling in black holes can
be interpreted as Brownian motion of information in the stretched
horizon with negative viscosity and negative temperature. For
alternative interpretations, see \cite{barb}-\cite{fish}.

At first, we shall illustrate our idea on a simple example and then
consider a more realistic model. We start with the Langevin equation
for a Brownian particle of mass $m$ immersed in a fluid with
temperature $T$ and viscosity $\zeta$
\begin{equation}
\label{3} m\ddot{\textbf{x}}=- \alpha
\dot{\textbf{x}}+\textbf{f}(t),
\end{equation}
where $\alpha$ is the frictional constant proportional to viscosity
(for a spherical particle with radius $r$, $\alpha =6\pi \zeta r$)
and $\textbf{f}(t)$ is a random force with zero average,
$\langle\textbf{f}(t)\rangle =0$. It can be shown (see any course on
stochastic processes) that the general expression for the
mean-square displacement of the particle in two spatial dimensions
is given by
\begin{equation}
\label{4} \langle x^{2}\rangle=\frac{4k_B
T}{\alpha}[t-\gamma^{-1}(1-e^{-\gamma t})],
\end{equation}
where $\gamma=\alpha/m$. For $\zeta > 0$ and $t\gg \gamma^{-1}$ we
get the famous result $\langle x^{2}\rangle = 4Dt$, where $D$ is the
diffusion coefficient, $D=k_B T/\alpha$. According to kinetic theory
$D\sim l_0 \bar{p}$, where $l_0$ is the mean free path and $\bar{p}$
is the average momentum of a molecule in thermal equilibrium. Then,
by using the fact that $\langle x^{2}\rangle \sim N l_0^{2}$ and
$\bar{p}^{2}\sim T$, we get $t_D T\sim N l_0 \bar{p}$, where $N$ is
the number of molecules and $t_D$ is the time of diffusion.
Following Seiko and Susskind, we assume that for ordinary local
systems the scrambling time is at least as long as the diffusion
time. When the de Broglie wavelength becomes of order $l_0$, we
enter into the regime of strongly correlated quantum fluids and we
have $t_{\rm scr} T\sim N $. If the total entropy of the fluid is of
order the number of molecules, we may also write $t_{\rm scr} T\sim
S $. This is the old result of Seiko and Susskind in two dimensions.

Consider now the case $\zeta < 0$ and $t \gg |\gamma^{-1}|$. In this
case we get
\begin{equation}
\label{5} \langle x^{2}\rangle = 4D\gamma ^{-1}e^{\gamma t},
\end{equation}
a completely new result. Since $|\gamma^{-1}|\sim l_0\sim T^{-1}$,
we get $t_{\rm scr} T\sim \log S $ in the regime of strongly
correlated quantum fluids. This is of the same form as (\ref{1}).
Therefore, Brownian motion in a fluid with negative viscosity can be
as fast as the scrambling in black hole. The reason for this is
that, due to internal instability, such a fluid behaves as an
amplifying medium. Amplification of a macroscopic degree of freedom,
such as a laser or maser field, or the motion of a Brownian
particle, is a phenomenon of enormous scientific and technological
importance. In classical thermodynamics, amplification is not
possible in thermal equilibrium at positive temperature. In
contrast, due to the fluctuation-dissipation theorems, amplification
is a natural phenomenon, and fully consistent with the second law of
thermodynamics, in a heat bath at \emph{negative temperature}.
Indeed, according to the fluctuation-dissipation theorem in the
Nyquist form \cite{reif}, the spectral density of the correlation
function for the random force $\textbf{f}(t)$ is, up to a numerical
factor, the product of the friction coefficient $\gamma$ and the
temperature $T$. It is always \emph{positive}. Therefore, if the
viscosity $\zeta$ is negative, then so is the temperature $T$. Thus,
the exponential growth of the mean free path with time in (\ref{5})
can be viewed as a result of Brownian motion in a fluid with
negative viscosity and negative temperature. A difficulty arises,
however, when we consider black holes. The Hawking temperature is
positive. Where does the negative temperature come from?

We start answering the question with an important note. The
Schwarzschild metric is featureless: the black holes have no hairs.
What happens if we get closer to the event horizon? In the
near-horizon approximation, the Euclidean metric of a black hole
takes the Rindler form
\begin{equation}
\label{6}ds_E\approx\rho^{2}d(g_H
t_E)^{2}+d\rho^{2}+\frac{1}{4g_H^{2}}d\Omega^{2},
\end{equation}
where $\rho$ is the proper distance from the horizon and $g_H$ is
the surface gravity, $g_H=1/4GM$. Since $t_E$ appears in the metric
only through its square, the classical solutions will be the same
metrics for positive and negative temperature $T_H=t_E^{-1}$. In
\cite{bra}, Braden \emph{et al}. considered the question of
determining the density of states of the microcanonical ensemble for
a black hole in a thermally isolated box containing a fixed amount
of energy. They showed (see also Louko and Whiting \cite{lou}), that
negative temperature arises in a natural and consistent way for a
singled-valued action of black-hole thermodynamics. However, their
treatment does not indicate how a state of gravitational field at
negative temperature should be prepared. We shall not look for the
negative temperature states of gravitational field in this essay.
Instead, we shall investigate one detailed example of the
stretched-horizon dynamics, which demonstrates explicitly the
appearance of negative temperatures in black holes. The argument
goes as follows. According to the membrane paradigm \cite{mem}, the
dynamical equations of the stretched horizon are equivalent to the
stretched-horizon fluid equations. In particular, for a
Schwarzschild black hole, the focusing (Raychaudhuri) equation for
the horizon expansion $\theta_H$ is identical to the energy
conservation for a fluid with surface energy density
$\Sigma_H=-\theta_H/8\pi G$, surface pressure $P_H=g_H/8\pi G$,
shear and bulk viscosity $\eta_H=1/16\pi G$ and $\zeta_H=-1/16\pi
G$, respectively. For a complete understanding of the horizon-fluid
dynamics we must also take into account the thermodynamical
properties of a black hole. In particular, we shall regard a small
patch of the horizon-fluid of area $\Delta A$ as endowed with a
temperature $T_H=g_H/2\pi$ and entropy $\Delta S_H=\Delta A/4G$.
Then the energy conservation for the patch, when rewritten using the
expression for $\Sigma_H$, $P_H$, $\eta_H$, $\zeta_H$, and
$\theta_H=(1/\Delta A)(d\Delta A/dt)$, becomes the heating equation
(without shear terms) \cite{pri}, \cite{red}
\begin{equation}
\label{7} \frac{d^{2}\Delta S_H}{dt^{2}}= 2\pi T_H \frac{d\Delta
S_H}{dt}-2\pi\Delta A (\zeta_H \theta_H ^{2}+\mathcal{F}_H),
\end{equation}
where $\mathcal{F}_H$ is the flux of entropy across the horizon from
the external universe. For small $\theta_H$, the equation reduces to
\begin{equation}
\label{8} \frac{d^{2}\Delta S_H}{dt^{2}}= 2\pi T_H \frac{d\Delta
S_H}{dt}-2\pi\Delta A \mathcal{F}_H.
\end{equation}
In this form, it is similar to the Langevin equation (\ref{3}) with
\emph{negative} friction coefficient $-2\pi T_H $ and random force
$2\pi\Delta A \mathcal{F}_H$. Indeed,  Bhattacharya and
Shankaranarayanan showed in \cite{bha1} and \cite{bha2} that
Raychaudhuri equation for the horizon expansion $\theta_H$ can be
viewed as a Langevin equation. They noticed that $\Delta S_H$
($\sqrt{\Delta A}$, in their work) increases exponentially with time
$\Delta S_H \sim e^{tT_H}$. However, they did not realize that this
increase is related to the \emph{negative} temperature $T_H$. In our
opinion, it is the negative temperature $T_H$ that causes the
exponential increase of $\Delta S_H$. Therefore, we conclude that
propagation of a small perturbation in the horizon-fluid can be
regarded as Brownian motion in a fluid at the negative temperature
$T_H$ and the time required for the perturbation to diffuse in the
stretched-horizon fluid is the same as the scrambling time
(\ref{1}).

It must be realized, however, that for all real system the states
with negative temperatures are, strictly speaking, not equilibrium
states but merely metastable states. As a consequence of this, we
can speak of negative temperature only in some conditional sense.
The concept of negative temperature has a still more conditional
sense in the theory of quantum amplifiers of electromagnetic waves,
lasers and masers. The point is that the laser (maser) medium is not
in a thermal state but in a steady state. Therefore, it is not in
thermal equilibrium and cannot have a temperature. The concept of
negative temperature is used in lasers (masers) only as a
conditional quantity characterizing the inversion population
condition. In this essay we consider the negative temperature of the
stretched horizon-fluid in the same sense.

In our opinion, this is a key to understanding of the mystery of
non-locality in the fast scrambling. The point is that the stretched
horizon with negative temperature (and viscosity) behaves as an
inverted or amplifying medium. As is well known
\cite{ora}-\cite{boy}, in such a medium, the occurrence of a
perturbation at a certain point is associated with the amplification
of a part of harmonics already available at this point due to the
system's instability rather than with the energy (and information)
transmission. For example, a Gaussian wave packet has wings that
extend from $-\infty$ to $+\infty$, so that the wave packet is
literally everywhere at all times. It turns out \cite{chi}, the
inverted medium temporarily borrows a part of its stored energy to
the forward tail of the wave packet in reshaping process that moves
the peak of the wave packet forward in time. Note that, in contrast
to the wave equation, the diffusion equation being only first order
in time, implies that perturbation must propagate with infinite
speed. However, at the fundamental level, there is no superluminal
communication. Moreover, we can even find a situation \cite{boy}, in
which the output wave packet leaves the inverted medium before the
peak of the input wave packet enters. This is very much like the
teleological behavior of the black hole's horizon. However, in this
case, there is no violation of causality. Since the
stretched-horizon fluid behaves as an inverted or amplifying medium,
we conclude that there is no non-locality in the fast scrambling
process. The stretched horizon mimics non-locality; we have only an
illusion.

It is possible that instability caused by the negative temperature
(and negative viscosity) is related with dynamical chaos. In
particular, it is believed \cite{mal}, \cite{kit} that the rate of
scrambling can be interpreted in terms of chaos as a Lyapunov
exponent $\lambda_L$, $t_{\rm scr} \sim \lambda_L^{-1}\log S$. In
this connection it may be noted that the Lyapunov exponent for the
spreading process was first introduced and determined in \cite{ro}.

The existence of a lower bound on the scrambling time seems to imply
that the relation (\ref{1}) has its origin in the uncertainty
relation. Therefore, we want to conclude this essay by demonstrating
how (\ref{1}) can be derived from the uncertainty relation. As we
know from quantum mechanics \cite{lan}, the relation $\Delta E
\Delta x \sim \dot{x}$ exists between the quantum uncertainties of
energy and of some quantity $x$, $\dot{x}$ being the classical rate
of change of $x$. Let us suppose that quantum uncertainty of $x$ is
of the same order as $x$, so that $\Delta x \sim x$. Then we have
$\Delta E x \sim \dot{x}$. It follows from this that
\begin{equation}
\label{9} t \Delta E \sim \log x.
\end{equation}
Let us relate all these quantities with a black hole. As is well
known, if a fiducial observer who is a distance of order the
Schwarzschild radius $R_S$ from the horizon drops a freely falling
quantum information into the black hole, it takes Schwarzschild time
of order $t\sim R_S \log R_S$ for the information to reach the
stretched horizon, and conversely it takes Schwarzschild time of
order $R_S \log R_S$ for quanta emitted from the stretch horizon to
reach the observer's detector. Indeed,
\begin{equation}
\label{10} t \sim
\int^{R_S+\delta}_{2R_S}\frac{dr}{\sqrt{1-\frac{R_S}{r}}}\simeq R_ S
\log R_S,
\end{equation}
where (restoring for a moment the Plank constant) $\delta \sim l_
P^{2}/R_S$; remember that the stretched horizon is located about one
Planck unit of proper distance above the event horizon. Let us now
identify $x$ with $R_S$. Since we have $2R_S$ in the lower limit of
(\ref{10}), we can put $\Delta R_S \sim R_S$. For a black hole
$\Delta E$ is of order the black hole's temperature, $\Delta E\sim
T$. Finally, we get the scrambling time (\ref{1}) derived from the
uncertainty relation.

\end{document}